\documentclass[leqn,final,10pt]{wlscirep}
\usepackage[utf8]{inputenc}
\usepackage[T1]{fontenc}
\usepackage{lineno,hyperref}
\usepackage{amsmath}
\usepackage[normalem]{ulem}
\usepackage{graphicx}
\usepackage[markup=underlined]{changes}
 \DeclareMathOperator{\tsinc}{tsinc}
\usepackage{setspace}

\usepackage[justification=justified,
   format=plain]{caption} 

\usepackage{outlines}
\usepackage{textgreek}
\usepackage{soul}
\usepackage{subfig}
\usepackage{gensymb}
\usepackage{amsmath} 
\usepackage{nccmath}
\usepackage{float}
\usepackage{graphicx}
\usepackage[utf8]{inputenc}
\usepackage{mathtools}
\usepackage{caption}
\usepackage{multirow}

    \DeclareCaptionFormat{myformat}{\hspace*{1cm}#1}
\title{Ortho-positronium Lifetime For Soft-tissue Classification}

\author[a]{Ashish V. Avachat${}^{*}$}
\author[c]{Kholod H. Mahmoud}
\author[c]{Anthony G. Leja }
\author[d,e]{Jiajie J. Xu}
\author[b]{Mark A. Anastasio}
\author[f]{Mayandi Sivaguru}
\author[c]{Angela Di Fulvio${}^{*}$}
\affil[a]{Department of Mechanical Engineering and Materials Science, University of Pittsburgh}
\affil[b]{Department of Bioengineering, University of Illinois Urbana-Champaign}
\affil[c]{Department of Nuclear, Plasma, and Radiological Engineering, University of Illinois Urbana-Champaign}
\affil[d]{Department of Clinical Medicine, University of Illinois Urbana-Champaign}
\affil[e]{Animal Care Program, University of Illinois Urbana-Champaign}
\affil[f]{Cytometry and Microscopy to Omics Facility, Roy J Carver Biotechnology Center, University of Illinois Urbana-Champaign}
\affil[*]{ashish.avachat@pitt.edu; difulvio@illinois.edu}


\begin{abstract}

The objective of this work is to showcase the ortho-positronium lifetime as a probe for soft-tissue characterization. 
We employed positron annihilation lifetime spectroscopy to experimentally measure the three components of the positron annihilation lifetime\textemdash para-positronium (p-Ps), positron, and ortho-positronium (o-Ps)\textemdash for three types of porcine, non-fixated soft tissues \textit{ex vivo}: adipose, hepatic, and muscle.   
Then, we benchmarked our measurements with X-ray phase-contrast imaging, which is the current state-of-the-art for soft-tissue analysis. 
We found that the o-Ps lifetime in adipose tissues  (\replaced{2.54$\pm$0.12}{ 3.82$\pm$0.20} ns) was approximately 20\% longer than in hepatic (\replaced{2.04$\pm$0.09}{ 3.11$\pm$0.34} ns) and muscle (\replaced{2.03$\pm$0.12}{ 3.02$\pm$0.11} ns) tissues.
In addition, the separation between the measurements for adipose tissue and the other tissues was better from o-Ps lifetime measurement than from X-ray phase-contrast imaging.
This experimental study proved that the o-Ps lifetime is a viable non-invasive probe for characterizing and classifying the different soft tissues.
Specifically, o-Ps lifetime as a soft-tissue characterization probe had a strong sensitivity to the lipid content that can be potentially implemented in commercial positron emission tomography scanners that feature list-mode data acquisition. 
\end{abstract}

\begin{document}

\flushbottom
\maketitle

\thispagestyle{empty}

\section*{Introduction}
    Positron Annihilation Lifetime Spectroscopy (PALS) is a well-established, non-destructive analysis technique, widely used in the realm of material science since the seventies  \cite{Gidley2006,Tao1972}. 
    PALS employs positrons and their meta-stable products as probes for material characterization. 
  \textcolor{black}{  In a material to be characterized, the positrons recombine with the electrons of this host material and annihilate by emitting two back-to-back 511-keV photons. 
    Before this annihilation, a positron can alternatively interact with an electron by forming a bound exotic state called positronium (Ps).  
    Such Ps can appear in two forms: para-positronium (p-Ps, with antiparallel spin) or ortho-positronium (o-Ps, with parallel spin). 
    Typically, a p-Ps annihilates by emitting two back-to-back 511-keV photons and an o-Ps annihilates through a three-quantum process\cite{Goworek2015}, emitting three photons with a total energy of 1022 keV. 
    O-Ps annihilation can also yield higher odd multiplets with a much lower probability \cite{DeBenedetti19561209}; one-quantum annihilation of free o-Ps is instead prohibited.
    The mean lifetime of o-Ps in a vacuum is approximately 142 ns \cite{Ore1949}, but it decreases to hundreds of picoseconds to a few nanoseconds in matter \cite{van2016asymmetric, gidley1999positronium} because the positron of the bound pair can also annihilate with an electron of opposite spin of the surrounding medium. 
    This process of o-Ps annihilation with a lattice electron, which results in 2-$\gamma$ emission, is referred to as the o-Ps \textit{pick-off}. 
    As a result, o-Ps in matter can either annihilate emitting two 511-keV gamma rays after \textit{pick-off} or through the three-quantum intrinsic decay process.  
     The likelihood of each of these processes occurring depends on the material properties of the host\textemdash primarily, the electron density and the average radius of micro-structural voids within the material. 
    The distribution of void radii quantifies the distribution of the sizes of empty spaces or defects within the structure of the host material \cite{Goworek2015}.  
   The lifetime of the o-Ps and, hence, the ratio between 3$\gamma$ and 2$\gamma$ emissions in a given material are proportional to the average void radius of the material \cite{Tao1972, Goworek2015}.}

\begin{figure}[ht]
      \centering
        \centering
        \includegraphics[width=6.5in,angle=0]{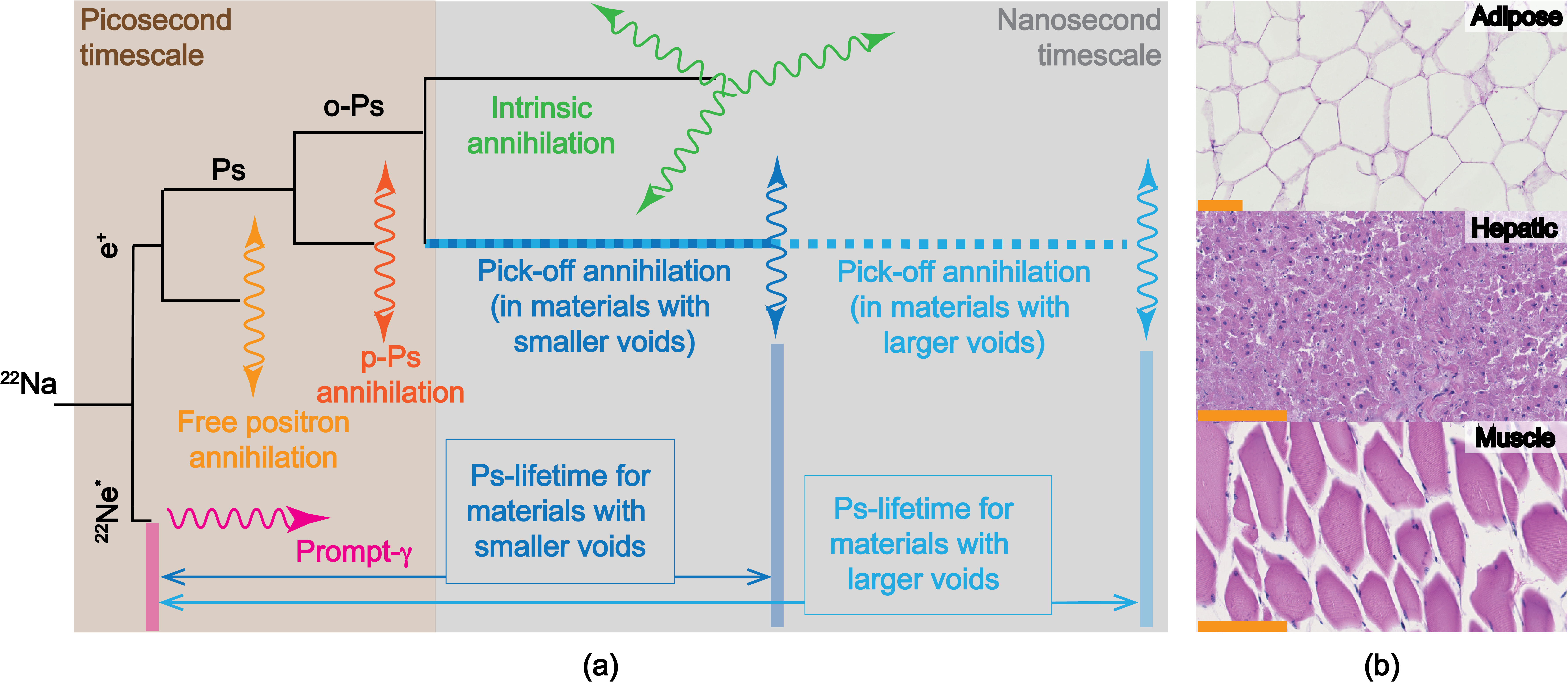}
    \caption{\textcolor{black}{\textbf{(a) Illustration of the physics of positronium transport.}
    ${}^{22}$Na undergoes ${\beta}^{+}$ decay (positron and neutrino emissions) and turns into an excited ${}^{22}{Ne}^{*}$, which in turn decays by emitting a prompt-$\gamma$ (1274.6 keV gamma emission).
    Hereafter, the emitted positron is referred to as free positron (FP).
    FP can either directly annihilate by 2$\gamma$ emission or can form a positronium (Ps); Ps can take a form of a para-positronium (p-Ps) or an ortho-positronium (o-Ps); p-Ps directly annihilates by 2$\gamma$ emission; and o-Ps can either annihilate by 2$\gamma$ emission (pick-off annihilation) or 3$\gamma$ emission.
    \textbf{(b) Histological photomicrographs of example soft tissues.} 
    Hematoxylin and eosin (H\&E) stained histological photomicrographs for adipose, hepatic, and muscle tissues. 
    Scale bars represent 100 $\mu$m.}}
    \label{fig:PALS_Introduction}
\end{figure}

    For instance, in materials with high electron density and small average void radius, o-Ps is more likely to annihilate via the two-quantum process rather than the three-quantum process.
    This phenomenon results in a shorter mean lifetime of o-Ps in materials with high electron density and a small average void radius.
    On the other hand, in materials with low electron density and large average void radius, the o-Ps can get trapped in the voids and is more likely to undergo the three-quantum process rather than annihilate via 2$\gamma$ emission. 
   Such conditions comprising low electron density and large average void radius result in a longer mean lifetime of o-Ps than in materials with high electron density and small average void radius.
        (Figure \ref{fig:PALS_Introduction}a illustrates the transport physics for a positron from a ${}^{22}$Na isotope.
    A comprehensive review of Ps interactions with matter can be found elsewhere \cite{DeBenedetti19561209}.)
    
    Under a nonrelativistic approximation, the lifetime of the positron in the bound pair ($\tau_{PS}$) decreases with the electron density ($N_{e}$) of the medium: $\tau_{PS} = 1/\pi r^{2}_o c N_{e}$, with $r_{o}$ being the classic electron radius ($2.80\times10^{-13}$ cm) and \textit{c} being the speed of light \cite{Tao1972}.
    The measurement of o-Ps lifetime ($\tau_{o-Ps}$) and the 3$\gamma$ annihilation fraction ($f_{3\gamma}$) \textemdash i.e., the ratio of the 3$\gamma$ annihilation events and the total number of annihilation events\textemdash can be used to characterize the density and average size of empty spaces in the medium. 
    Equation \ref{eq:lifetime2} applies to porous materials and shows that the o-Ps lifetime in the medium ($\tau_{o-Ps-medium}$) increases with the 3$\gamma$ annihilation fraction \cite{Goworek2015,jaksia1}: 
    \begin{ceqn}
    \begin{equation}
      \tau_{o-Ps-medium} =\left( f_{3\gamma}-\frac{\left( -\frac{4}{3}P_{o-Ps} \right) }{370} \right) \left(\frac{\tau_{o-Ps-vacuum}}{P_{o-Ps}}\right). \label{eq:lifetime2}
    \end{equation}
    \end{ceqn}
    In Equation \ref{eq:lifetime2}, $P_{o-Ps}$ is the probability of o-Ps formation in the material, $\tau_{o-Ps-vacuum}$ is the o-Ps lifetime in vacuum, and the constant 370 is the computed ratio of cross sections of 2$\gamma$ annihilation and 3$\gamma$ annihilation  ($\sigma_{2\gamma}/\sigma_{3\gamma} \approx 370$) for particles in free relative motion, as calculated by Ore and Powell using the time-dependent perturbation theory \cite{Ore1949}.  
    The o-Ps lifetime is also directly related to the average void radius, \textit{R} (Equation \ref{eq:lifetime}\cite{Goworek2015,jaksia1}), as:
  \begin{ceqn}
  \begin{equation}
       \tau_{o-Ps-medium}=\frac{1}{\lambda_b}\left(1-\frac{R}{R-\Delta}+\frac{1}{2\pi}sin\left(\frac{2 \pi R}{R+\Delta}\right)\right)^{-1}. \label{eq:lifetime}
  \end{equation}
 \end{ceqn}
    In Equation \ref{eq:lifetime}, $\lambda_b$ is the spin-averaged positronium annihilation rate ($\lambda_b \approx 1/4 \lambda_{S} = 2 $ ns$^{-1}$) and $\Delta$ is an experimentally measured parameter\cite{Goworek2015}.
    This theory shows that the positron lifetime directly decreases with an increase in electron density, but o-Ps lifetime  ($\tau_{o-Ps-medium}$) increases with an increase in the average void size in the medium (Equation \ref{eq:lifetime}). 
    Importantly, $\tau_{o-Ps-medium}$ and, hence, the average void size can be estimated by measuring the fraction of 3$\gamma$-decay events (Equation \ref{eq:lifetime2}). 
    The potential use of o-Ps lifetime to characterize some biological tissues has been recently explored, wherein the viability of using Ps and o-Ps lifetimes to analyze tissue properties, such as oxygen concentration and tissue porosity, was reported \cite{Jasinska2017,Stepanov2021,Shibuya2020,Axpe_2014}.
    The presence of voids, air pockets, and defects results in a higher probability for the o-Ps to be trapped inside them, which in turn results in longer mean o-Ps-lifetime and a higher $f_{3\gamma}$, than the void-absent conditions.
    The measurement of the mean o-Ps-lifetime provides an indication of the average size of the voids.
    In the biology domain, $f_{3\gamma}$ also decreases with the density of free radicals, which have unpaired electrons that easily undergo pick-off \cite{Jasinska2017}. 
    Using these mechanisms, some promising results have been previously reported in terms of discriminating between healthy and tumor tissues \cite{Moskal2019a,Zgardzinska2020,Stepanov2021,Moskal2021,Axpe_2014,Stepien_2021,Karimi_2023,Moskal_2023}, but a generalized characterization framework for discriminating different types of soft tissues has not yet been reported. 
    
    In this paper, we report the measured Ps lifetimes in porcine, non-fixated soft tissue samples: adipose, hepatic, and muscle (see Figure \ref{fig:PALS_Introduction}b for histological photomicrographs of example soft tissues).a
    These experimental measurements agreed with the previously discussed analytical relationships. 
    To establish PALS as a probe for soft tissue analysis and to benchmark the sensitivity of the PALS to the subtle changes in the soft tissues, we compared its performance with the X-ray phase-contrast computed tomography (XPC-CT), which is the current state-of-the-art in non-invasive soft tissue analysis \cite{romell2021virtual}. 
    
\section*{Experimental results and discussion}
We measured the PALS spectra for three types of porcine, non-fixated,
soft tissues \textit{ex vivo}: adipose, hepatic, and muscle. 
These spectra, the objects of our analysis, consisted of the distribution of the differences in detection times between the 1.27 MeV prompt-$\gamma$ from $^{22}$Na decay chain and the positron products\textemdash two- and three-quantum annihilation gamma rays.

\subsection*{Analysis of the positronium annihilation lifetime spectra
}
\begin{figure}[htbp]
    \centering
  \includegraphics[width=\linewidth]{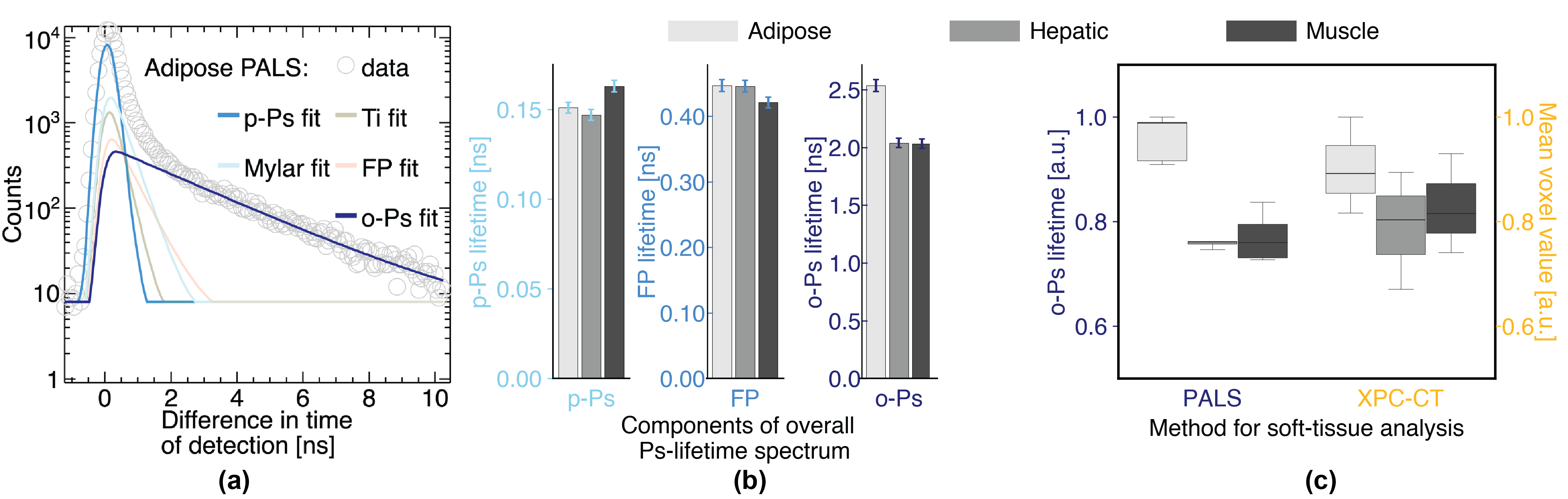}
    \caption{\textcolor{black}{\textbf{Results of PALS analysis and its benchmarking with XPC-CT.} 
    (\textit{a}) The five components Ps-lifetime\textemdash p-Ps, \added{fixed component for titanium (Ti) foil encapsulation of the source (248 ps\cite{Mcguire_2006}),}
    \replaced{}{fixed source component (382 ps),} \added{fixed component for Mylar foil that was used to wrap the tissue samples (382 ps\cite{Ning_2017}),} free-positron (FP), \replaced{}{fixed holder component (1.63 ns),} and o-Ps lifetimes\textemdash can be decomposed from the measured PALS spectrum (example PALS spectrum and its decomposed components for adipose tissues in \textit{a}). 
    (\textit{b}) The comparison of three non-fixed components of the PALS spectrum for the three soft tissue types with 5 repeats for each type (n = 5) showed that o-Ps was the most sensitive out of the three components for discerning the subtle changes between the different types of soft-tissue. 
    Single standard-deviation error bars are added to the plot.
    (\textit{c}) Comparison of PALS as a method for soft-tissue analysis with the current state-of-the-art, XPC-CT.
    PALS probe measurement\textemdash o-Ps lifetime\textemdash showed less variations across the samples than the mean voxel value measurement through XPC-CT. 
    The adipose tissue was significantly more discernible from the hepatic and the muscle tissues using mean o-Ps lifetime compared to using mean voxel values in XPC-CT. 
    (For the box-and-whiskers plot in \textit{c}, the PALS measurements and XPC-CT mean voxel values were normalized to their respective maximums for presentation purposes.)}
    }
    \label{fig:PALS_Results}
\end{figure}

    We resolved five components of lifetimes from the PALS spectra (Figure \ref{fig:PALS_Results}a). 
    Each of these components characterizes one specific process of the positron interactions \cite{FANG2019162507}. 
    (Two of these five components were fixed during the analysis of the PALS spectra and are not discussed in this section\added{, but are detailed in Methods section}: \replaced{one corresponding to the 5.10-$\mu m$-thick titanium foils that encapsulated the $^{22}$Na source, 248 ps\cite{Mcguire_2006}, and another corresponding to the 2.5-$\mu m$-thick Mylar foils that were wrapped around the tissue samples}{ one corresponding to the Kapton source, 382 ps\cite{Dulski2017} with 10\% intensity, and another corresponding to the holder material, 1.63 ns}).
    From the fastest to the slowest, the first component corresponds to p-Ps decay ($\tau_1$, $I_1$), the second to the annihilation of free positrons ($\tau_2$, $I_2$), and the third to o-Ps self-decay ($\tau_3$, $I_3$) \cite{FANG2019162507}.
    The PALS spectrum, \textit{f(t)}, can be modeled as a convolution of the exponential decay functions corresponding to each process and the detector time resolution function (Equation \ref{eq:1}), as follows:
\begin{ceqn}
    \begin{equation}
    f(t) = \sum_{i=1}^3 \frac{I_i}{\tau_i} \mathrm{e}^{-\frac{t}{\tau_i}} * \frac{\mathrm{e}^{-\frac{t^2}{2\sigma^2}}}{\sqrt{2\pi\sigma^2}} 
    =  \sum_{i=1}^3 \frac{I_i}{2\tau_i} \mathrm{e}^{\frac{\sigma^2-2\tau_it}{2\tau_{i}^{2}}} \mathrm{erfc}\left(\frac{\sigma^2-\tau_it}{\sqrt{2}\sigma\tau_{i}}\right).
    \label{eq:1}
    \end{equation}
\end{ceqn}

    In Equation (\ref{eq:1}), $\tau_{i}$ and $I_{i}$ are the lifetime and intensity of the $i$-th component, and $\sigma$ represents the time resolution of the detection system (198.3 $\pm$ 0.8 ps\cite{FANG2019162507}). 
    This excellent time resolution was achieved through the optimization of the experimental PALS setup and timing algorithms described in our previous work \cite{FANG2019162507}.
    Two of the decay constants of the overall Ps-lifetime\textemdash the ones that correspond to the p-Ps decay ($\tau_1$) and to the free-positron (FP) annihilation ($\tau_2$)\textemdash were comparable in the three types of tissues (Figure \ref{fig:PALS_Results}b). 
    On the other hand, the o-Ps lifetime ($\tau_3$) component of the PALS spectrum showed a significant difference (about 20\%) for the adipose tissue compared to that of hepatic and muscle tissues (Figure \ref{fig:PALS_Results}b; the mean o-Ps lifetimes w/ mean intensities for the adipose, hepatic, and muscle tissues with five repeats for each tissue type: \replaced{2.54$\pm$0.12~ns w/ 18.60$\pm$2.90\%, 2.04$\pm$0.09~ns w/ 12.92$\pm$2.36\%, 2.03$\pm$0.12~ns w/ 14.47$\pm$2.89\%}{ 3.82$\pm$0.20~ns w/ 18.20$\pm$2.86\%, 3.11$\pm$0.34~ns w/ 12.68$\pm$3.56\%, and 3.02$\pm$0.11~ns w/ 13.46$\pm$1.62\%}, respectively). 
    \replaced{Here, the uncertainties}{
    The uncertainty} in the o-Ps lifetime\replaced{s were}{ was} calculated as the sampled uncertainties of five samples of the same tissue utilizing the confidence interval method \cite{bipm2020guide}.

    An interesting comparison with previous work by Moskal et al. is worthy of noting here: the difference between the o-Ps lifetimes of adipose and the cardiac myxoma for humans was found to be about 30\% \cite{Moskal_2023}. 
    Although our counterpart of this difference was about 20\% and not an exact match with that presented in Moskal et al.\cite{Moskal_2023} study, the trend of o-Ps lifetime of adipose tissue being higher than that of a myocyte-based tissue agreed with each other in these two studies. 
    This disparity between the exact values of the said differences can  be associated to 1) the two studies have two different species (humans tissues for Moskal et al.\cite{Moskal_2023} and porcine tissues for ours); and 2) the myocyte-based tissue counterparts of the two studies were of different types (anomalous cardiac muscle tissues for Moskal et al.\cite{Moskal_2023} and healthy skeletal muscle tissues for ours).
    However, a more comprehensive study involving o-Ps lifetime measurements of different species and biophysical analysis of positronium interaction is required to confidently draw inferences on such cross-species correlations and on explicitly describing the possible underlying mechanisms for the different o-Ps lifetimes.
    (One of such possible underlying mechanisms is the oxygen concentration \cite{Moskal2021,2022_Moyo}).

\subsection*{PALS analysis for soft-tissue discrimination and its benchmarking with the current state-of-the-art}    
    We found that the adipose tissues can easily be discriminated from the other two tissues by thresholding the mean o-Ps lifetime (Figure \ref{fig:PALS_Results}c).
    The difference between the mean o-Ps lifetime for hepatic and muscle tissues was subtle, yet discernible (Figure \ref{fig:PALS_Results}b and c).
    Based on the previously discussed positronium transport physics, the significant difference in mean o-Ps lifetimes in adipose tissues and in hepatic and muscle tissues can be attributed primarily to the different structural porosity. 

    In order to experimentally corroborate these observations of mean o-Ps lifetimes, we compared our o-Ps lifetime measurements with the current state-of-the-art in soft tissue analysis\cite{romell2021virtual,Donato_2022,Mareike2016,Mareike2018,Frohn2020}: X-ray phase-contrast computed tomography (XPC-CT). 
    The analysis of 3D image data from XPC-CT informed us about the relative contrast that is currently achievable and can be used for discriminating the selected soft tissues. 
    The benchmarking with XPC-CT showed that the mean o-Ps lifetime followed a similar trend as the mean voxel value from XPC-CT (Figure \ref{fig:PALS_Results}c).
    The mean XPC-CT voxel values for soft tissue samples depend mainly on their mass densities and effective atomic number (z$_{eff}$). 
    XPC-CT results showed that the X-ray attenuation coefficients of the analyzed muscle and hepatic tissues are comparable, on the other hand, the adipose tissues exhibit a lower attenuation coefficient, owing to its lower density and z$_{eff}$, as expected. 
    Similarly, o-Ps lifetime is longer in adipose tissues than in hepatic and muscle tissues.
    However, the relative difference between the signal from adipose tissues and that from the other tissues is notably higher in PALS than XPC-CT. 
    This result indicates that relatively fewer o-Ps collisions\textemdash that result in ``pick-off"\textemdash occur in adipose tissues than in hepatic and muscle tissues. 
    This effect is likely due to the porous structure of adipocytes filled with low density lipids.    
    In addition, we found that the inter-sample variations in the mean o-Ps lifetimes for the selected soft tissues was significantly lower than that of mean voxel value (Figure \ref{fig:PALS_Results}c). 
    Combined, the findings from this comparison\textemdash one, the higher difference between the mean o-Ps lifetime of adipose tissue and that of hepatic and muscle tissues and, two, the lower inter-sample variations\textemdash show that it is easier to identify a threshold to discriminate adipose tissues from the hepatic and muscle tissues by employing PALS analysis than the current state-of-the-art, XPC-CT.

    It should be noted that our XPC-CT imaging system employed an extended propagation distance for enhancing the phase effects in a laboratory setting. 
    Although such propagation-based XPC-CT is widely used for soft-tissue analysis as a virtual histology tool\cite{romell2021virtual,Donato_2022,Mareike2016,Mareike2018,Frohn2020}, it is only one of the forms of acquiring phase-contrast. 
    One of the other forms of acquiring phase-contrast is by employing grating-interferometry. 
    The effect of larger average void radii, which our propagation-based XPC-CT system was unable to resolve, might be captured using grating-based XPC-CT, because such  system will provide an image that maps X-ray scattering \cite{Massimi2022,Polikarpov2023}. 
    A comparison between mean o-Ps lifetime  with mean voxel values for diffraction or scattering images from grating-based XPC-CT for soft tissue analysis will be a valuable next step. 

\section*{Conclusion}
    We analyzed three types of porcine, non-fixated soft tissues\textemdash adipose, hepatic, and muscle\textemdash using positron annihilation lifetime spectroscopy (PALS).
    The PALS spectra included contributions from three annihilation lifetimes: p-Ps, free-positron, and o-Ps. 
    We found that the o-Ps lifetime in adipose tissue was, on average, 20\% longer than in hepatic and muscle tissues.
    Although it is known that mean o-Ps lifetime increases with porosity in inorganic materials; in biological tissues, mean o-Ps lifetime is the convolution of several phenomena, such as oxygen and free radical concentration, in addition to the average void radius. 
    The oxygen partial pressure was alike in all the measured samples, and no free radicals were induced in the tissues. 
    Therefore, the significantly higher response of adipose tissue\textemdash in terms of o-Ps lifetime, when compared to hepatic and muscle tissue\textemdash  can likely be ascribed to a higher average void radius, which was not resolved by the benchmarking XPC-CT imaging.
    Given the recent surge in positronium imaging within the health sciences domain \cite{Moskal2019a,Shibuya_2022,Moskal2021,Moskal2022,Qi2022,}, the impact of this result in this domain is that mean o-Ps lifetime can be used as a non-invasive probe for detecting and quantifying lipid content in soft tissues or organs.
    
\section*{Methods}

\textcolor{black}{    We acquired porcine tissue samples of adipose, hepatic, and muscle types from the Meat Science Laboratory of the University of Illinois Urbana-Champaign (five samples of each type from five castrated males; 24 weeks old, PIC 800 sires x PIC Camborough dams). 
    Each tissue sample was analyzed, consecutively, by PALS and XPC-CT in a non-fixated state.}
    
 \subsection*{PALS: measurements and analysis}    
     \begin{figure}[h!]
        \centering
            \begin{tabular}{c}
                \subfloat[]{
                    \centering \includegraphics[height=1.75in]{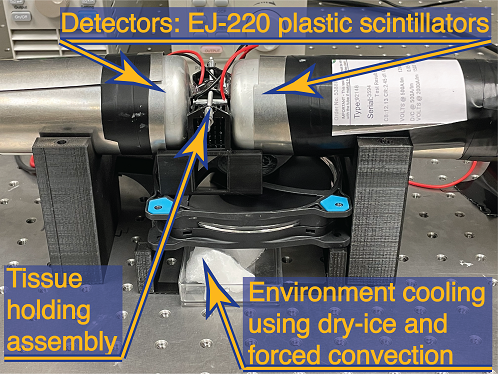}
                }
                \subfloat[]{
                    \centering \includegraphics[height=1.75in]{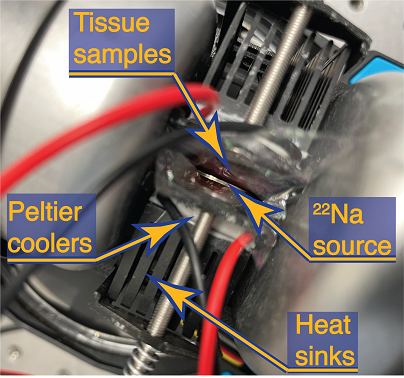}
                }
                \subfloat[]{
                    \centering \includegraphics[height=1.75in]{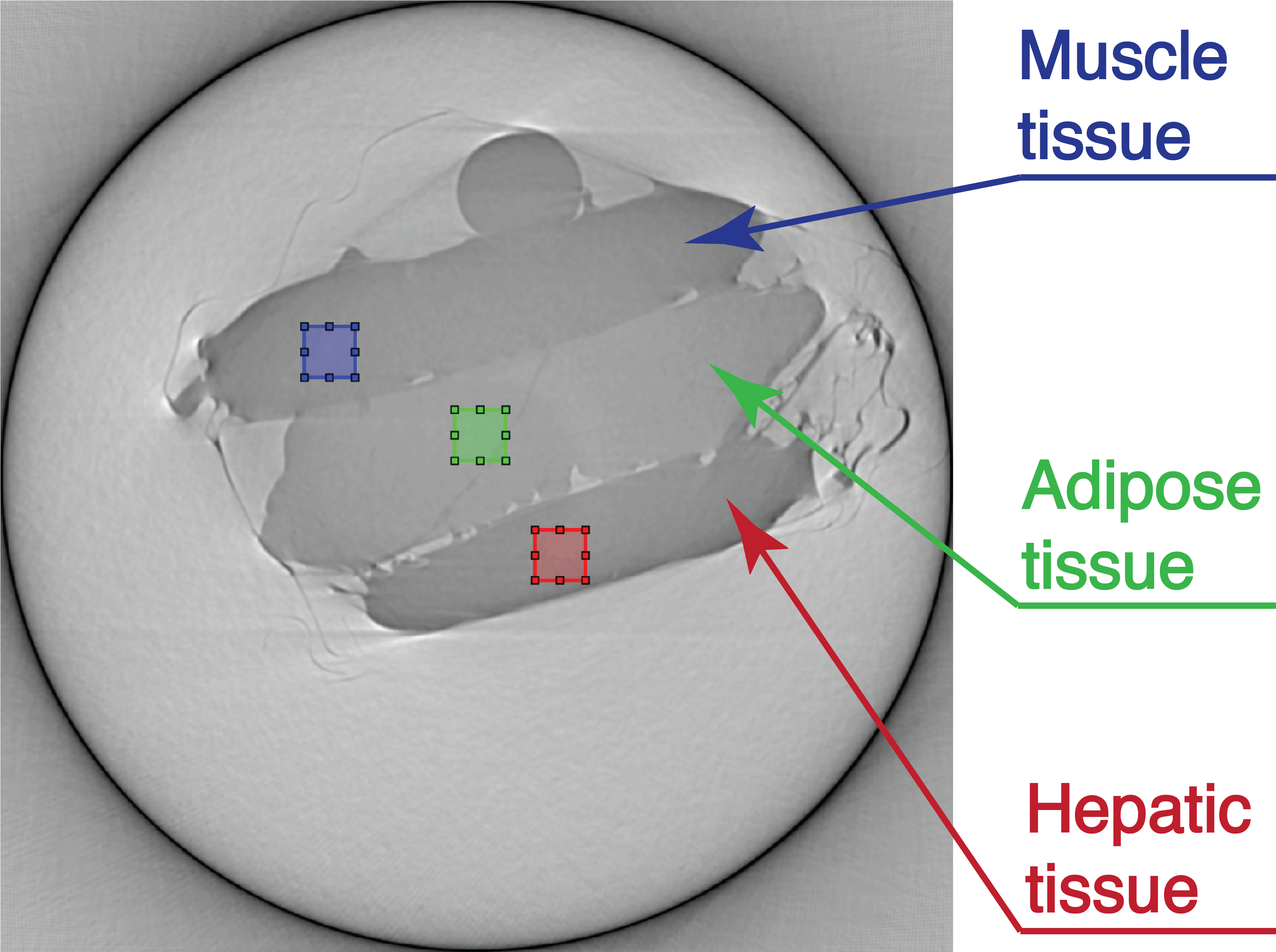}
                }
                \\
                \subfloat[]{
                    \centering \includegraphics[height=1.4in]{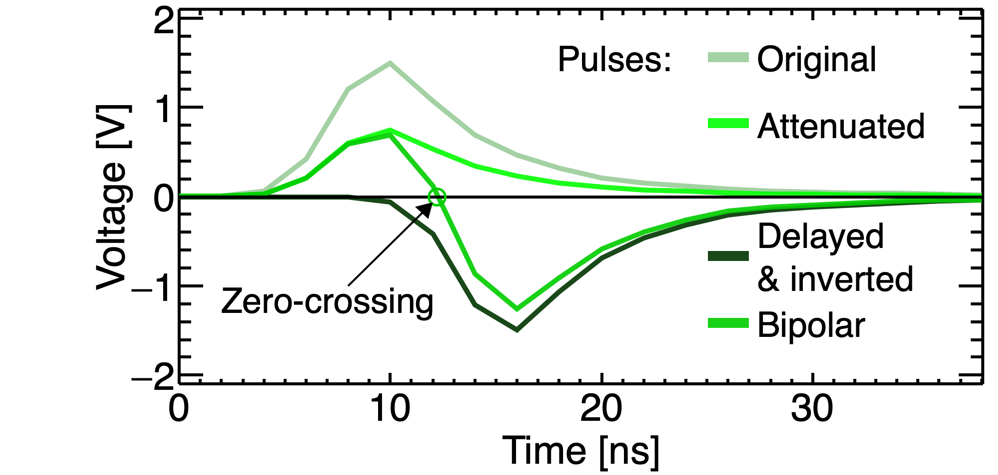}
                }
                
                \subfloat[]{
                    \centering \includegraphics[height=1.4in]{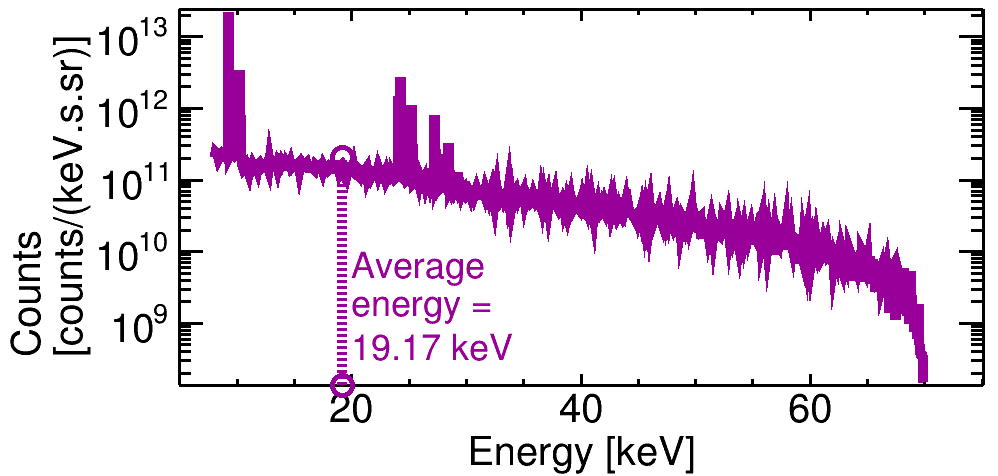}
                }

            \end{tabular}
            \caption{\textbf{Methods for PALS analysis and its benchmarking with XPC-CT.} 
            (\textit{a}) The PALS measurement setup: two scintillation detectors with tissue samples in the middle irradiated by the $^{22}$Na source.
            (\textit{b}) $^{22}$Na source was placed between two tissue samples that were kept cooled by a Peltier-cell based system. 
            \added{The tissue samples were wrapped with Mylar foils of 2.5 $\mu m$ thickness to avoid these tissue samples from touching the $^{22}$Na source.}
            (\textit{c}) An example tomographic slice of XPC-CT images. The blue, green, and red boxes represent a 2D slice of the 3D regions-of-interest corresponding to the adipose, hepatic, and muscle tissues, respectively, in the given example XPC-CT slice.
            (\textit{d}) Sample fast pulse acquired by the organic scintillation detector, (1) its attenuated version by a factor \textit{F}, (2) the original pulse delayed by $\mathrm{\Delta}$~ns and inverted, and the (3) bipolar pulse obtained by subtracting (2) from the original pulse. The zero-crossing point corresponds to the pulse time stamp.
            (\textit{e}) Energy spectrum and its average energy for 70 kVp X-ray beam from the liquid-metal-jet-based X-ray source used by XPC-CT system. 
            }
        \label{fig:PALS_Methods}
    \end{figure}

    We extracted the p-Ps, free positron (FP), and o-Ps lifetime components from the PALS spectra for porcine, non-fixated adipose, hepatic, and muscle tissues.
    The PALS spectra were acquired by using a $^{22}$Na source and by recording the distribution of the differences in detection times between the 1.27 MeV  prompt decay gamma ray and the positron products\textemdash annihilation and decay gamma rays.
    \added{The positron emitting ${^{22}}$Na source used for the presented measurements was a POSN series positron source from Eckert \& Ziegler, POSN-22\cite{EZSource}, with initial activity of 10-$\mu$Ci on February 1$^{st}$ 2023; and the presented measurements were completed between June 20$^{th}$ and 23$^{rd}$ 2023. 
    In this PSON-22 $^{22}$Na source, the positron emitting active diameter of 9.53 mm is encapsulated between two titanium foils (5.10 $\mu m$ thickness), which is then supported by two 250 $\mu m$ thick titanium disks (19.10 mm outer diameter and 9.53 mm inner diameter) sealed by electron beam welding to form a brim for the source.}
    In the experimental setup for PALS measurement (Figures  \ref{fig:PALS_Methods}a and \ref{fig:PALS_Methods}b), \replaced{this source}{ a 10-$\mu$Ci activity ${^{22}}$Na source, the positron emitter,} was placed between two samples of a given tissue type (one type at a time: adipose, hepatic, and muscle) with five repeats of each type to capture inter-sample statistical variations. 
    By such sandwiching of the positron emitter with two samples, we maximized the tissue interaction and reduced the overall data acquisition time.  
    \added{Each tissue sample was wrapped with a 2.5 $\mu m$-thick Mylar foils to avoid the tissues samples from directly touching the $^{22}$Na source.}
    This assembly of source and two tissue samples was secured using a 3D-printed holder made of acrylonitrile butadiene styrene (ABS, Figure \ref{fig:PALS_Methods}a). 
    The o-Ps lifetime for the holder material was measured before and after each tissue type for consistency (mean o-Ps lifetime: 1.63 ns).
    \added{But, during the PALS measurements for the tissues samples, the probability of positrons reaching the holder was expected to be negligible because of the high thickness (5 mm) of the tissue samples that completely covered the active area of the ${^{22}}$Na source, axially, and because of about 4.8 mm radial width of the titanium brim of the source that can stop majority of radially emitted positrons.}
    \replaced{}{Then two of the five components were fixed when decomposing the PALS spectra\textemdash one for Kapton source to 382 ps\cite{Dulski2017} with 10\% intensity and another for this holder to 1.63 ns}.

    \replaced{}{This $^{22}$Na source was encapsulated in thin Mylar foils, which ensured that most of the positrons emitted by the source reached the tissue samples.} 
    The $^{22}$Na emits a positron, and the daughter nucleus de-excites in about 3~ps by emitting a 1.27 MeV gamma ray.
    Therefore, the emission of the 1.27~MeV was considered as a timestamp for the $\beta^+$ decay (positron emission time).
    \textcolor{black}{During the irradiation, the approximate 1~cm$\times$1~cm$\times$0.5~cm tissue samples were kept at approximately 9$^\circ$C by employing an environment cooling system based on Peltier coolers, dry-ice, and forced convection (Figures \ref{fig:PALS_Methods}a and \ref{fig:PALS_Methods}b).
    The temperature of the sample was measured using a Fluke-62 MAX Infrared Thermometer.
    Two fast EJ-220 plastic organic scintillators of 5.08~cm length and 5.08~cm in diameter coupled to 9214B ETEL photomultiplier vacuum tubes (PMT) were used to detect gamma rays in coincidence. 
    We selected EJ-220 because of its fast light response; previously, we reported a time resolution of 183$\pm$0.8 ps with this measurement system\cite{FANG2019162507}.} 
    We used a high-voltage power supply (DT5533EN, CAEN Technologies) to power the PMTs. 
    The detected pulses were directly digitized by a 14-bit 500 MSps digitizer (DT5730, CAEN Technologies) and transferred to a workstation as full waveforms in binary format using the acquisition software CoMPASS \cite{compass}. 

    Each pulse, detected upon gamma-ray interaction in the scintillator, was associated with a time stamp. 
    Accurate pulse timing is crucial to reconstruct the PALS spectrum successfully \cite{FANG2019162507}. 
    Therefore, we developed and optimized a constant fraction discrimination (CFD) algorithm to accurately find each pulse onset time \cite{repo}. 
    Digital CFD provides accurate timing and is superior to other timing algorithms (e.g., analog CFD or leading edge) in terms of time resolution when used for processing the organic scintillation pulses \cite{Steinberger2019185}. 
    Pulses were digitized at a 500 MSps sampling frequency using the DT5730 digitizer. 
    We interpolated these digitized voltage values, one every 2~ns, to densely sample the rising edge, which typically lasts less than 10~ns. 
    Each pulse can be modeled as the convolution between the sampled voltage values and a terminated $sinc$ function ($tsinc$) \cite{warburton2017new}.
    This model is detector-invariant and satisfies the Nyquist condition \cite{shannon1998communication, FANG2019162507}. 
    The $tsinc$ function is a $sinc$ function modulated by a Gaussian function and is needed for time-finite, non-periodic signals. 
    Equation \ref{eq:5} is the \textit{k}-th linearly interpolated sample obtained after dividing the measurement sampling time into \textit{N} intervals inside a window of width $L$, where, $g_s(j)$ is the \textit{j}-th sampling value of the input waveform: 
\begin{equation}
\begin{aligned}
    G(j, k) &= \sum_{i=0}^{L-1}g_s\left(j-i\right)\tsinc\left(i+\frac{k}{N}\right) +g_s\left(j+1+i\right)\tsinc\left(i+1-\frac{k}{N}\right). 
    \label{eq:5}
\end{aligned}
\end{equation}
    After pulse interpolation, the time stamp was determined by the CFD algorithm:
\begin{equation}
    \mathrm{CFD}\left(l\right) = F\times g_s\left(l\right) - g_s\left(l-\Delta\right),
    \label{eq:6}
\end{equation}
    where, $g_s(l)$ is $l$-th sample of the interpolated pulse, with $l=j \times N+k$, CFD$(l)$ is a bipolar pulse with the zero-crossing point being the time stamp, the attenuation factor $F$ (0-1) and time shift $\mathrm{\Delta}$ are two data processing parameters that depend on the scintillator type. 
    The pulse timing was obtained as the pulse zero-crossing time after subtracting the original pulse inverted and delayed by $\mathrm{\Delta}$ nanoseconds from an attenuated version of the original pulse (Figure \ref{fig:PALS_Methods}d). 
    For our experimental setup, the optimum values for $F$ and $\mathrm{\Delta}$ are 0.4~ns and 4~ns, respectively \cite{FANG2019162507}.

    The annihilation and self-decay of the Ps in the tissue samples resulted in the emission of two or three gamma rays. 
    The emission of two gamma rays in coincidence (the prompt 1.27~MeV followed by a lower energy gamma ray) was detected by the scintillator pair by applying a coincidence time gate of 200~ns, after determining the pulse detection times.
    The PALS spectra were created by histogramming the time interval between the arrival times of the 1.27~MeV decay gamma ray and of lower-energy annihilation or decay gamma rays. 
    Each sample was measured for 180 minutes, obtaining approximately 130,000 time-coincidence counts contributing to the PALS distribution in each run.
    The inherent low energy resolution of organic scintillation detectors does not allow us to discriminate 511~keV from lower energy gamma rays from 3$\gamma$ self-decay, but their use in this study was mainly motivated by their fast response.
    As a result, the discrimination of the positron lifetime components was merely based on the cumulative PALS spectra.
    PALS spectra are well described by the linear combination of different components: the fast decay of p-Ps, the slow decay of FP, and the delayed self-decay of o-Ps (Equation \ref{eq:1}). 
    We used the software LT10 \cite{Kansy2011} to decompose five components: p-Ps component, \added{fixed component for the titanium (Ti) foil encapsulation of the $^{22}$Na source (248 ps\cite{Mcguire_2006}),} \replaced{}{fixed source component (382 ps with 10\% intensity \cite{Dulski2017}),} \added{fixed component for the Mylar foils that were used for wrapping the tissue samples (382 ps\cite{Ning_2017}),} FP component, \replaced{}{fixed holder component (1.63 ns for ABS, measured by us),} and o-Ps component for tissue sample.
    LT10 is one of the most popular software for positron lifetime spectra analysis, which was developed by Kansy and Giebel \cite{GIEBEL2012122, KANSY1996235}. It relies on the deconvolution of the experimental data into the model's exponential functions followed by a nonlinear fitting procedure to find the model’s parameters that best fit the measured PALS distribution \cite{KANSY1996235}. The user can either select one of the predefined LT10 models to fit the data or implement a custom one. We used the model described in Equation \ref{eq:1} to fit our data, with  $\tau_{1}$, $\tau_{2}$, $\tau_{3}$, $I_{1}$, $I_{2}$, and $I_{3}$ as parameters with a lower bound to zero. The time resolution of our measurement system is also needed in the PALS model ($\sigma$) and was previously determined experimentally to be 198.3 $\pm$ 0.8 ps   \cite{FANG2019162507}; therefore, $\sigma$ is a constant in the LT10 model. Further details on the LT10 software can be found elsewhere \cite{Kansy2011,GIEBEL2012122}.
    The analysis using LT10 also provided the fitting uncertainties associated with the derived p-Ps, FP, and o-Ps lifetimes in each tissue sample. 
    The error bars in Figure \ref{fig:PALS_Results}b account for the propagation of the 1-standard-deviation uncertainties associated with the fitted parameters of the lifetime distribution measured on multiple samples from the same tissue batch.
    We calibrated this PALS measurement and analysis procedure by employing a Certified Reference Material (CRM)\textemdash polycarbonate (National Metrology Institute of Japan NMIJ CRM 5602-a)\textemdash and a previously characterized quartz sample\cite{FANG2019162507}. 
    The polycarbonate samples were 15$\times$15$\times$1.5 mm$^3$ in size, and the quartz samples were 10$\times$10$\times$1 mm$^3$ each. 
    We measured the o-Ps lifetime to be 2.07 $\pm$ 0.05 ns and 1.38 $\pm$ 0.05 ns for polycarbonate and quartz, respectively. 
    These results agreed, within two standard deviations, with the one provided by the vendor for the polycarbonate CRM (2.10 $\pm$ 0.05 ns\cite{CRM}) and with the previous measurement for quartz (1.34 $\pm$ 0.05 ns\cite{FANG2019162507}), respectively.

  \subsection*{X-ray phase-contrast imaging: measurements and analysis} 
        The X-ray phase-contrast imaging of the selected non-fixated soft tissues was performed using the propagation-based X-ray phase-contrast CT (XPC-CT) system of the Computational X-ray Imaging Science Laboratory at the Beckman Institute of the University of Illinois Urbana-Champaign \cite{Deshpande_2023}. 
        This XPC-CT system included: a liquid-metal-jet-based X-ray source (MetalJet D2, Excillum) operated at 70 kVp, 130~W power, and about 14 $\mu m$ focal spot size; a CsI(Tl) scintillator-based X-ray imager with a pixel pitch of 13 $\mu m$, active area of about ${54}^{2}$ ${mm}^{2}$, and 2x2 binning (4k $\times$ 4k X-Ray GSENSE SCMOS, Photonic Science); and an object manipulation system to rotate the imaging samples.
        A propagation distance (object-to-detector distance) of approximately 2.15 m was utilized to enhance the phase effects.
        This propagation distance, paired with the source-to-object distance of approximately 1.85 m, resulted in a magnification of 2.16 and Fresnel number of approximately 2.82 for an average X-ray energy of 19.17 keV for the 70 kVp X-ray energy spectrum of our liquid-metal-jet-based X-ray source (Figure \ref{fig:PALS_Methods}e).
        A set of 720 projections was acquired to fully scan the imaging object (0\degree ~to 360\degree) with an exposure time of about 2.4 s per projection. 
        Paganin's phase-retrieval toolbox, ANKAPhase, was used to recover the phase map for each projection \cite{Weitkamp2011}.
        An in-house implementation of the Feldkamp, Davis, and Kress (FDK) algorithm was used to reconstruct the tomographic volume from the phase-contrast enhanced projections\cite{Feldkamp1984}.
        For XPC-CT scanning, a PMMA rod and the three soft tissue samples were stacked together, with each wrapped in an ultra-thin Mylar sheets. 
        (PMMA rod was added to the assembly for being able to quantify and normalize the mean voxel values between air- and water-equivalent attenuation.) 
        This assembly was taped together tightly such that the field-of-view (FOV= 25 mm) had the following sequence (top to bottom in Figure \ref{fig:PALS_Methods}c): PMMA rod, muscle sample, adipose sample, and hepatic sample. 
        The mean voxel values for each sample (5 samples of each tissue type) were calculated by randomly inserting 3D regions-of-interest (${100}^{3}$ voxels) with ten repeats (Figure \ref{fig:PALS_Methods}c).
        
\subsection*{Histology: measurements}   
    One of each soft tissue types were fixed under pressure and vacuum with 10\% formalin after PALS measurement and XPC-CT imaging.
    Then fixed samples were embedded in paraffin, and ten slices of 8-10 micron thickness were cut using a microtome from each block sample (1 cm$^{3}$).
    Two sections were extracted from each slice. 
    Out of the ten slides for each type of sample, five were stained using Mason's Trichrome and the remaining five using hematoxylin and eosin. 
    Microscopy was performed using a Hamamatsu Nanozoomer using a 20x 0.75 NA Olympus objective. 
    
\section*{Data Availability}
The datasets used and/or analysed during the current study are available from the corresponding author on reasonable request.
\bibliography{Ref}

\begin{thebibliography}{10}
\urlstyle{rm}
\expandafter\ifx\csname url\endcsname\relax
  \def\url#1{\texttt{#1}}\fi
\expandafter\ifx\csname urlprefix\endcsname\relax\def\urlprefix{URL }\fi
\expandafter\ifx\csname doiprefix\endcsname\relax\def\doiprefix{DOI: }\fi
\providecommand{\bibinfo}[2]{#2}
\providecommand{\eprint}[2][]{\url{#2}}

\bibitem{Gidley2006}
\bibinfo{author}{Gidley, D.~W.}, \bibinfo{author}{Peng, H.~G.} \& \bibinfo{author}{Vallery, R.~S.}
\newblock \bibinfo{journal}{\bibinfo{title}{Positron annihilation as a method to characterize porous materials}}.
\newblock {\emph{\JournalTitle{Annual Review of Materials Research}}} \textbf{\bibinfo{volume}{36}}, \bibinfo{pages}{49--79}, \doiprefix\url{10.1146/annurev.matsci.36.111904.135144} (\bibinfo{year}{2006}).

\bibitem{Tao1972}
\bibinfo{author}{Tao, S.~J.}
\newblock \bibinfo{journal}{\bibinfo{title}{Positronium annihilation in molecular substances}}.
\newblock {\emph{\JournalTitle{The Journal of Chemical Physics}}} \textbf{\bibinfo{volume}{56}}, \bibinfo{pages}{5499--5510}, \doiprefix\url{10.1063/1.1677067} (\bibinfo{year}{1972}).

\bibitem{Goworek2015}
\bibinfo{author}{Goworek, T.}
\newblock \bibinfo{journal}{\bibinfo{title}{Positronium as a probe of small free volumes in crystals, polymers and porous media}}.
\newblock {\emph{\JournalTitle{Annales UMCS, Chemia}}} \textbf{\bibinfo{volume}{69}}, \bibinfo{pages}{1--110}, \doiprefix\url{10.2478/umcschem-2013-0012} (\bibinfo{year}{2015}).

\bibitem{DeBenedetti19561209}
\bibinfo{author}{DeBenedetti, S.}
\newblock \bibinfo{journal}{\bibinfo{title}{New atoms - positronium and mesonic atoms}}.
\newblock {\emph{\JournalTitle{Il Nuovo Cimento}}} \textbf{\bibinfo{volume}{4}}, \bibinfo{pages}{1209--1270}, \doiprefix\url{10.1007/BF02744346} (\bibinfo{year}{1956}).

\bibitem{Ore1949}
\bibinfo{author}{Ore, A.} \& \bibinfo{author}{Powell, J.~L.}
\newblock \bibinfo{title}{Three-photon annihilation of an electron-positron pair} (\bibinfo{year}{1949}).

\bibitem{van2016asymmetric}
\bibinfo{author}{Van~Horn, J.~D.}, \bibinfo{author}{Wu, F.}, \bibinfo{author}{Corsiglia, G.} \& \bibinfo{author}{Jean, Y.~C.}
\newblock \bibinfo{title}{Asymmetric positron interactions with chiral quartz crystals?}
\newblock In \emph{\bibinfo{booktitle}{Defect and Diffusion Forum}}, vol. \bibinfo{volume}{373}, \bibinfo{pages}{221--226}, \doiprefix\url{10.4028/www.scientific.net/DDF.373.221} (\bibinfo{organization}{Trans Tech Publ}, \bibinfo{year}{2016}).

\bibitem{gidley1999positronium}
\bibinfo{author}{Gidley, D.} \emph{et~al.}
\newblock \bibinfo{journal}{\bibinfo{title}{Positronium annihilation in mesoporous thin films}}.
\newblock {\emph{\JournalTitle{Physical Review B}}} \textbf{\bibinfo{volume}{60}}, \bibinfo{pages}{R5157}, \doiprefix\url{10.1103/PhysRevB.60.R5157} (\bibinfo{year}{1999}).

\bibitem{jaksia1}
\bibinfo{author}{Jasi\'{n}ska, B.} \& \bibinfo{author}{Moskal, P.}
\newblock \bibinfo{journal}{\bibinfo{title}{A new pet diagnostic indicator based on the ratio of 3$\gamma$/2$\gamma$ positron annihilation}}.
\newblock {\emph{\JournalTitle{Acta Physica Polonica B}}} \textbf{\bibinfo{volume}{48}}, \bibinfo{pages}{1577--1582}, \doiprefix\url{10.5506/APhysPolB.48.1577} (\bibinfo{year}{2017}).

\bibitem{Jasinska2017}
\bibinfo{author}{Jasi\'{n}ska, B.} \emph{et~al.}
\newblock \bibinfo{title}{Human tissues investigation using pals technique}.
\newblock vol.~\bibinfo{volume}{48}, \bibinfo{pages}{1737--1747}, \doiprefix\url{10.5506/APhysPolB.48.1737} (\bibinfo{publisher}{Jagellonian University}, \bibinfo{year}{2017}).

\bibitem{Stepanov2021}
\bibinfo{author}{Stepanov, S.~V.}, \bibinfo{author}{Byakov, V.~M.} \& \bibinfo{author}{Stepanov, P.~S.}
\newblock \bibinfo{journal}{\bibinfo{title}{Positronium in biosystems and medicine: A new approach to tumor diagnostics based on correlation between oxygenation of tissues and lifetime of the positronium atom}}.
\newblock {\emph{\JournalTitle{Physics of Wave Phenomena}}} \textbf{\bibinfo{volume}{29}}, \bibinfo{pages}{174--179}, \doiprefix\url{10.3103/S1541308X21020138} (\bibinfo{year}{2021}).

\bibitem{Shibuya2020}
\bibinfo{author}{Shibuya, K.}, \bibinfo{author}{Saito, H.}, \bibinfo{author}{Nishikido, F.}, \bibinfo{author}{Takahashi, M.} \& \bibinfo{author}{Yamaya, T.}
\newblock \bibinfo{journal}{\bibinfo{title}{Oxygen sensing ability of positronium atom for tumor hypoxia imaging}}.
\newblock {\emph{\JournalTitle{Communications Physics}}} \textbf{\bibinfo{volume}{3}}, \doiprefix\url{10.1038/s42005-020-00440-z} (\bibinfo{year}{2020}).

\bibitem{Axpe_2014}
\bibinfo{author}{Axpe, E.} \emph{et~al.}
\newblock \bibinfo{journal}{\bibinfo{title}{Detection of atomic scale changes in the free volume void size of three-dimensional colorectal cancer cell culture using positron annihilation lifetime spectroscopy}}.
\newblock {\emph{\JournalTitle{PLOS ONE}}} \textbf{\bibinfo{volume}{9}}, \bibinfo{pages}{1--5}, \doiprefix\url{10.1371/journal.pone.0083838} (\bibinfo{year}{2014}).

\bibitem{Moskal2019a}
\bibinfo{author}{Moskal, P.}, \bibinfo{author}{Jasi{\'{n}}ska, B.}, \bibinfo{author}{St{\c{e}}pie{\'{n}}, E.} \& \bibinfo{author}{Bass, S.~D.}
\newblock \bibinfo{journal}{\bibinfo{title}{{Positronium in medicine and biology}}}.
\newblock {\emph{\JournalTitle{Nature Reviews Physics}}} \textbf{\bibinfo{volume}{1}}, \bibinfo{pages}{527--529}, \doiprefix\url{10.1038/s42254-019-0078-7} (\bibinfo{year}{2019}).

\bibitem{Zgardzinska2020}
\bibinfo{author}{Zgardzi{\'{n}}ska, B.} \emph{et~al.}
\newblock \bibinfo{journal}{\bibinfo{title}{{Studies on healthy and neoplastic tissues using positron annihilation lifetime spectroscopy and focused histopathological imaging}}}.
\newblock {\emph{\JournalTitle{Scientific Reports}}} \textbf{\bibinfo{volume}{10}}, \bibinfo{pages}{1--10}, \doiprefix\url{10.1038/s41598-020-68727-3} (\bibinfo{year}{2020}).

\bibitem{Moskal2021}
\bibinfo{author}{Moskal, P.} \emph{et~al.}
\newblock \bibinfo{journal}{\bibinfo{title}{{Positronium imaging with the novel multiphoton PET scanner}}}.
\newblock {\emph{\JournalTitle{Science Advances}}} \textbf{\bibinfo{volume}{7}}, \bibinfo{pages}{1--10}, \doiprefix\url{10.1126/sciadv.abh4394} (\bibinfo{year}{2021}).

\bibitem{Stepien_2021}
\bibinfo{author}{St{\c{e}}pie{\'{n}}, E.} \emph{et~al.}
\newblock \bibinfo{journal}{\bibinfo{title}{{Positronium life-time as a new approach for cardiac masses imaging}}}.
\newblock {\emph{\JournalTitle{European Heart Journal}}} \textbf{\bibinfo{volume}{42}}, \bibinfo{pages}{ehab724.3279}, \doiprefix\url{10.1093/eurheartj/ehab724.3279} (\bibinfo{year}{2021}).

\bibitem{Karimi_2023}
\bibinfo{author}{Karimi, H.}, \bibinfo{author}{Moskal, P.}, \bibinfo{author}{Zak, A.} \& \bibinfo{author}{St{\c{e}}pie{\'{n}}, E.~L.}
\newblock \bibinfo{journal}{\bibinfo{title}{3{D} melanoma spheroid model for the development of positronium biomarkers}}.
\newblock {\emph{\JournalTitle{Scientific Reports}}} \textbf{\bibinfo{volume}{13}}, \bibinfo{pages}{7648}, \doiprefix\url{10.1038/s41598-023-34571-4} (\bibinfo{year}{2023}).

\bibitem{Moskal_2023}
\bibinfo{author}{Moskal, P.} \emph{et~al.}
\newblock \bibinfo{journal}{\bibinfo{title}{Developing a novel positronium biomarker for cardiac myxoma imaging}}.
\newblock {\emph{\JournalTitle{EJNMMI Physics}}} \textbf{\bibinfo{volume}{10}}, \bibinfo{pages}{22}, \doiprefix\url{10.1186/s40658-023-00543-w} (\bibinfo{year}{2023}).

\bibitem{romell2021virtual}
\bibinfo{author}{Romell, J.}
\newblock \emph{\bibinfo{title}{Virtual histology by laboratory x-ray phase-contrast tomography}}.
\newblock Ph.D. thesis, \bibinfo{school}{KTH Royal Institute of Technology} (\bibinfo{year}{2021}).

\bibitem{FANG2019162507}
\bibinfo{author}{Fang, M.}, \bibinfo{author}{Bartholomew, N.} \& \bibinfo{author}{{Di Fulvio}, A.}
\newblock \bibinfo{journal}{\bibinfo{title}{Positron annihilation lifetime spectroscopy using fast scintillators and digital electronics}}.
\newblock {\emph{\JournalTitle{Nuclear Instruments and Methods in Physics Research Section A: Accelerators, Spectrometers, Detectors and Associated Equipment}}} \textbf{\bibinfo{volume}{943}}, \bibinfo{pages}{162507}, \doiprefix\url{https://doi.org/10.1016/j.nima.2019.162507} (\bibinfo{year}{2019}).

\bibitem{Mcguire_2006}
\bibinfo{author}{McGuire, S.} \& \bibinfo{author}{Keeble, D.~J.}
\newblock \bibinfo{journal}{\bibinfo{title}{Positron lifetimes of polycrystalline metals: {A} positron source correction study}}.
\newblock {\emph{\JournalTitle{Journal of Applied Physics}}} \textbf{\bibinfo{volume}{100}}, \bibinfo{pages}{103504}, \doiprefix\url{10.1063/1.2384794} (\bibinfo{year}{2006}).

\bibitem{bipm2020guide}
\bibinfo{author}{BIPM, I.}, \bibinfo{author}{IFCC, I.}, \bibinfo{author}{ISO, I.} \& \bibinfo{author}{IUPAP, O.}
\newblock \bibinfo{journal}{\bibinfo{title}{Guide to the expression of uncertainty in measurement—part 6: Developing and using measurement models}}.
\newblock {\emph{\JournalTitle{Joint Committee for Guides in Metrology, GUM-6}}}  (\bibinfo{year}{2020}).

\bibitem{2022_Moyo}
\bibinfo{author}{Moyo, S.}, \bibinfo{author}{MOSKAL, P.} \& \bibinfo{author}{St{\c{e}}pie{\'{n}}, E.~l.}
\newblock \bibinfo{journal}{\bibinfo{title}{Feasibility study of positronium application for blood clots structural characteristics}}.
\newblock {\emph{\JournalTitle{Bio-Algorithms and Med-Systems}}} \textbf{\bibinfo{volume}{18}}, \bibinfo{pages}{163--167}, \doiprefix\url{10.2478/bioal-2022-0087} (\bibinfo{year}{2022}).

\bibitem{Ning_2017}
\bibinfo{author}{Ning, X.} \emph{et~al.}
\newblock \bibinfo{journal}{\bibinfo{title}{Modification of source contribution in {PALS} by simulation using {Geant4} code}}.
\newblock {\emph{\JournalTitle{Nuclear Instruments and Methods in Physics Research Section B: Beam Interactions with Materials and Atoms}}} \textbf{\bibinfo{volume}{397}}, \bibinfo{pages}{75--81}, \doiprefix\url{10.1016/j.nimb.2017.02.038} (\bibinfo{year}{2017}).

\bibitem{Donato_2022}
\bibinfo{author}{Donato, S.} \emph{et~al.}
\newblock \bibinfo{journal}{\bibinfo{title}{Optimization of pixel size and propagation distance in x-ray phase-contrast virtual histology}}.
\newblock {\emph{\JournalTitle{Journal of Instrumentation}}} \textbf{\bibinfo{volume}{17}}, \bibinfo{pages}{C05021}, \doiprefix\url{10.1088/1748-0221/17/05/C05021} (\bibinfo{year}{2022}).

\bibitem{Mareike2016}
\bibinfo{author}{T{\"o}pperwien, M.}, \bibinfo{author}{Krenkel, M.}, \bibinfo{author}{Quade, F.} \& \bibinfo{author}{Salditt, T.}
\newblock \bibinfo{title}{{Laboratory-based x-ray phase-contrast tomography enables 3D virtual histology}}.
\newblock In \bibinfo{editor}{Khounsary, A.~M.} \& \bibinfo{editor}{van Dorssen, G.~E.} (eds.) \emph{\bibinfo{booktitle}{Advances in Laboratory-based X-Ray Sources, Optics, and Applications V}}, vol. \bibinfo{volume}{9964}, \bibinfo{pages}{99640I}, \doiprefix\url{10.1117/12.2246460}. \bibinfo{organization}{International Society for Optics and Photonics} (\bibinfo{publisher}{SPIE}, \bibinfo{year}{2016}).

\bibitem{Mareike2018}
\bibinfo{author}{Töpperwien, M.}, \bibinfo{author}{van~der Meer, F.}, \bibinfo{author}{Stadelmann, C.} \& \bibinfo{author}{Salditt, T.}
\newblock \bibinfo{journal}{\bibinfo{title}{Three-dimensional virtual histology of human cerebellum by x-ray phase-contrast tomography}}.
\newblock {\emph{\JournalTitle{Proceedings of the National Academy of Sciences}}} \textbf{\bibinfo{volume}{115}}, \bibinfo{pages}{6940--6945}, \doiprefix\url{10.1073/pnas.1801678115} (\bibinfo{year}{2018}).
\newblock \eprint{https://www.pnas.org/doi/pdf/10.1073/pnas.1801678115}.

\bibitem{Frohn2020}
\bibinfo{author}{Frohn, J.} \emph{et~al.}
\newblock \bibinfo{journal}{\bibinfo{title}{{3D virtual histology of human pancreatic tissue by multiscale phase-contrast X-ray tomography}}}.
\newblock {\emph{\JournalTitle{Journal of Synchrotron Radiation}}} \textbf{\bibinfo{volume}{27}}, \bibinfo{pages}{1707--1719}, \doiprefix\url{10.1107/S1600577520011327} (\bibinfo{year}{2020}).

\bibitem{Massimi2022}
\bibinfo{author}{Massimi, L.} \emph{et~al.}
\newblock \bibinfo{journal}{\bibinfo{title}{Volumetric high-resolution x-ray phase-contrast virtual histology of breast specimens with a compact laboratory system}}.
\newblock {\emph{\JournalTitle{IEEE Transactions on Medical Imaging}}} \textbf{\bibinfo{volume}{41}}, \bibinfo{pages}{1188--1195}, \doiprefix\url{10.1109/TMI.2021.3137964} (\bibinfo{year}{2022}).

\bibitem{Polikarpov2023}
\bibinfo{author}{Polikarpov, M.} \emph{et~al.}
\newblock \bibinfo{journal}{\bibinfo{title}{Towards virtual histology with x-ray grating interferometry}}.
\newblock {\emph{\JournalTitle{Scientific Reports}}} \textbf{\bibinfo{volume}{13}}, \bibinfo{pages}{9049}, \doiprefix\url{10.1038/s41598-023-35854-6} (\bibinfo{year}{2023}).

\bibitem{Shibuya_2022}
\bibinfo{author}{Shibuya, K.}, \bibinfo{author}{Saito, H.}, \bibinfo{author}{Tashima, H.} \& \bibinfo{author}{Yamaya, T.}
\newblock \bibinfo{journal}{\bibinfo{title}{Using inverse laplace transform in positronium lifetime imaging}}.
\newblock {\emph{\JournalTitle{Physics in Medicine and Biology}}} \textbf{\bibinfo{volume}{67}}, \bibinfo{pages}{025009}, \doiprefix\url{10.1088/1361-6560/ac499b} (\bibinfo{year}{2022}).

\bibitem{Moskal2022}
\bibinfo{author}{Moskal, P.} \& \bibinfo{author}{St{\c{e}}pie{\'{n}}, E.~L.}
\newblock \bibinfo{journal}{\bibinfo{title}{Perspectives on translation of positronium imaging into clinics}}.
\newblock {\emph{\JournalTitle{Frontiers in Physics}}} \textbf{\bibinfo{volume}{10}}, \doiprefix\url{10.3389/fphy.2022.969806} (\bibinfo{year}{2022}).

\bibitem{Qi2022}
\bibinfo{author}{Qi, J.} \& \bibinfo{author}{Huang, B.}
\newblock \bibinfo{journal}{\bibinfo{title}{Positronium lifetime image reconstruction for tof pet}}.
\newblock {\emph{\JournalTitle{IEEE Transactions on Medical Imaging}}} \textbf{\bibinfo{volume}{41}}, \bibinfo{pages}{2848--2855}, \doiprefix\url{10.1109/TMI.2022.3174561} (\bibinfo{year}{2022}).

\bibitem{}
\bibinfo{author}{Jasi\'{n}ska, B.} \& \bibinfo{author}{Moskal, P.}
\newblock \bibinfo{title}{A new pet diagnostic indicator based on the ratio of 3γ/2γ positron annihilation}.
\newblock vol.~\bibinfo{volume}{48}, \bibinfo{pages}{1577--1582}, \doiprefix\url{10.5506/APhysPolB.48.1577} (\bibinfo{publisher}{Jagellonian University}, \bibinfo{year}{2017}).

\bibitem{EZSource}
\bibinfo{title}{Eckert \& {Ziegler} {Reference} and {Calibration} {Sources}}.
\newblock \bibinfo{howpublished}{\url{https://www.ezag.com/fileadmin/user_upload/isotopes/isotopes/Isotrak/isotrak-pdf/Product_literature/EZIPL/EZIP_catalogue_reference_and_calibration_sources.pdf}}.
\newblock \bibinfo{note}{Accessed: 2024-08-06}.

\bibitem{compass}
\bibinfo{author}{{CAEN Technologies}}.
\newblock \bibinfo{title}{Compass multiparametric daq software for physics applications}.
\newblock \bibinfo{howpublished}{\url{https://www.caen.it/products/compass/}} (\bibinfo{year}{2020}).

\bibitem{repo}
\bibinfo{title}{{PALS} pulse processing program based on the root framework}.
\newblock \bibinfo{howpublished}{\url{https://gitlab.engr.illinois.edu/nml/pals}}.
\newblock \bibinfo{note}{Accessed: 2022-02-10}.

\bibitem{Steinberger2019185}
\bibinfo{author}{Steinberger, W.}, \bibinfo{author}{Ruch, M.}, \bibinfo{author}{{Di Fulvio}, A.}, \bibinfo{author}{Clarke, S.} \& \bibinfo{author}{Pozzi, S.}
\newblock \bibinfo{journal}{\bibinfo{title}{Timing performance of organic scintillators coupled to silicon photomultipliers}}.
\newblock {\emph{\JournalTitle{Nuclear Instruments and Methods in Physics Research Section A: Accelerators, Spectrometers, Detectors and Associated Equipment}}} \textbf{\bibinfo{volume}{922}}, \bibinfo{pages}{185 – 192}, \doiprefix\url{10.1016/j.nima.2018.11.099} (\bibinfo{year}{2019}).

\bibitem{warburton2017new}
\bibinfo{author}{Warburton, W.~K.} \& \bibinfo{author}{Hennig, W.}
\newblock \bibinfo{journal}{\bibinfo{title}{New algorithms for improved digital pulse arrival timing with sub-gsps adcs}}.
\newblock {\emph{\JournalTitle{IEEE Transactions on Nuclear Science}}} \textbf{\bibinfo{volume}{64}}, \bibinfo{pages}{2938--2950}, \doiprefix\url{10.1109/TNS.2017.2766074} (\bibinfo{year}{2017}).

\bibitem{shannon1998communication}
\bibinfo{author}{Shannon, C.~E.}
\newblock \bibinfo{journal}{\bibinfo{title}{Communication in the presence of noise}}.
\newblock {\emph{\JournalTitle{Proceedings of the IEEE}}} \textbf{\bibinfo{volume}{86}}, \bibinfo{pages}{447--457}, \doiprefix\url{10.1109/JPROC.1998.659497} (\bibinfo{year}{1998}).

\bibitem{Kansy2011}
\bibinfo{author}{Kansy, J.} \& \bibinfo{author}{Giebel, D.}
\newblock \bibinfo{title}{Study of defect structure with new software for numerical analysis of pal spectra}.
\newblock vol. \bibinfo{volume}{265}, \doiprefix\url{10.1088/1742-6596/265/1/012030} (\bibinfo{publisher}{Institute of Physics Publishing}, \bibinfo{year}{2011}).

\bibitem{GIEBEL2012122}
\bibinfo{author}{Giebel, D.} \& \bibinfo{author}{Kansy, J.}
\newblock \bibinfo{journal}{\bibinfo{title}{Lt10 program for solving basic problems connected with defect detection}}.
\newblock {\emph{\JournalTitle{Physics Procedia}}} \textbf{\bibinfo{volume}{35}}, \bibinfo{pages}{122--127}, \doiprefix\url{https://doi.org/10.1016/j.phpro.2012.06.022} (\bibinfo{year}{2012}).
\newblock \bibinfo{note}{Positron Studies of Defects 2011}.

\bibitem{KANSY1996235}
\bibinfo{author}{Kansy, J.}
\newblock \bibinfo{journal}{\bibinfo{title}{Microcomputer program for analysis of positron annihilation lifetime spectra}}.
\newblock {\emph{\JournalTitle{Nuclear Instruments and Methods in Physics Research Section A: Accelerators, Spectrometers, Detectors and Associated Equipment}}} \textbf{\bibinfo{volume}{374}}, \bibinfo{pages}{235--244}, \doiprefix\url{https://doi.org/10.1016/0168-9002(96)00075-7} (\bibinfo{year}{1996}).

\bibitem{CRM}
\bibinfo{title}{Reference {Material} {Certificate}: {NMIJ} {CRM} 56028-a {\textbar} {Polycarbonate} for {Positron} {Hole}-size {Measurement}}.

\bibitem{Deshpande_2023}
\bibinfo{author}{Deshpande, R.}, \bibinfo{author}{Avachat, A.}, \bibinfo{author}{Brooks, F.~J.} \& \bibinfo{author}{Anastasio, M.~A.}
\newblock \bibinfo{journal}{\bibinfo{title}{Investigating the robustness of a deep learning-based method for quantitative phase retrieval from propagation-based x-ray phase contrast measurements under laboratory conditions}}.
\newblock {\emph{\JournalTitle{Physics in Medicine and Biology}}} \textbf{\bibinfo{volume}{68}}, \bibinfo{pages}{085005}, \doiprefix\url{10.1088/1361-6560/acc2aa} (\bibinfo{year}{2023}).

\bibitem{Weitkamp2011}
\bibinfo{author}{Weitkamp, T.}, \bibinfo{author}{Haas, D.}, \bibinfo{author}{Wegrzynek, D.} \& \bibinfo{author}{Rack, A.}
\newblock \bibinfo{journal}{\bibinfo{title}{{ANKAphase: Software for single-distance phase retrieval from inline X-ray phase-contrast radiographs}}}.
\newblock {\emph{\JournalTitle{Journal of Synchrotron Radiation}}} \textbf{\bibinfo{volume}{18}}, \bibinfo{pages}{617--629}, \doiprefix\url{10.1107/S0909049511002895} (\bibinfo{year}{2011}).

\bibitem{Feldkamp1984}
\bibinfo{author}{Feldkamp, L.~A.}, \bibinfo{author}{Davis, L.~C.} \& \bibinfo{author}{Kress, J.~W.}
\newblock \bibinfo{journal}{\bibinfo{title}{{Practical cone-beam algorithm}}}.
\newblock {\emph{\JournalTitle{J. Opt. Soc. Am. A}}} \textbf{\bibinfo{volume}{1}}, \bibinfo{pages}{612--619}, \doiprefix\url{10.1364/JOSAA.1.000612} (\bibinfo{year}{1984}).
\newblock \eprint{1312.3977}.

\end{thebibliography}

\section*{Acknowledgements}

    This work was funded in-part by the Nuclear Regulatory Commission, United States award number 31310018K0002. 
    We thank Karen Doty for preparing the tissue samples for the histology analysis and Ming Fang for his help to setup the PALS experiment and use the LT10 software.
    We also thank Anna Dilger and Bailey Harsh of Meat Science Laboratory for helping in acquiring the tissue samples.

\section*{Author contributions statement}

    A. Avachat and A. Di Fulvio conceived the experiment(s);  A. Avachat, A. Leja, K. Mahmoud, J. Xu, M. Sivaguru and A. Di Fulvio conducted the experiment(s) and processed the data, A. Avachat, A. Leja, K. Mahmoud, and A. Di Fulvio wrote the manuscript; A. Avachat, and A. Di Fulvio analyzed the results. All authors reviewed the manuscript.

\end{document}